\documentclass[12pt,preprint]{aastex}
\shorttitle{Spin-Down of Young Pulsars}
\shortauthors{Alpar et al.}
\begin{document}
\title{Pulsar Spindown by a Fall-Back Disk and the P-\.{P} Diagram}
\author{M. Ali Alpar}
\affil{Faculty of Engineering and Natural Science, Sabanc\i\ University,\\
    Orhanl\i\--Tuzla, Istanbul 81474, Turkey}
\author{A\c{s}k\i n Ankay and Efe Yazgan}
\affil{Department of Physics, Middle East Technical University, 
   Ankara 06531, Turkey}
\begin{abstract}
     Neutron stars may be surrounded by fall-back disks formed
from supernova core-collapse. If the disk circumscribes the
light-cylinder, the neutron star will be an active radio  pulsar 
spinning down under the propeller spin-down torque applied by the disk 
as well as the usual magnetic dipole radiation torque. Evolution 
across the $P-\dot{P}$ diagram is very rapid when  pulsar spin-down is 
dominated by the propeller torque. This explains the distribution of 
pulsars in the $P-\dot{P}$ diagram.
\end{abstract}


\section{Introduction}
Efforts to understand the newly
identified classes of neutron stars, in particular anomalous
X-ray pulsars (AXPs - Mereghetti 1999)  and soft gamma-ray repeaters
(SGRs - Woods et al. 1999) have followed two
avenues. Magnetar models, involving neutron star dipole magnetic
fields $B \sim10^{14}-10^{15}$ G, above the quantum critical field
$B_c=m^2c^3/e\hbar= 4.4 \times 10^{13}$ G, were
advanced to explain the mechanism and energetics of soft
gamma-ray repeaters (Thompson \& Duncan 1995).  
Alternative models propose to explain the new classes of neutron stars 
in terms of conventional $B\sim10^{12}$ G fields. These models 
involve accretion or propeller (Illarionov \& Sunyaev 1975)
torques from an accretion disk surrounding the isolated neutron stars
(Alpar 1999, 2001; Chatterjee, Hernquist \& Narayan 2000). Alpar (1999,
2001) argued that 
radio pulsars, dim thermal neutron 
stars (DTNs; Treves et al. 2000), AXPs and radio quiet neutron stars 
(RQNSs; Chakrabarty et al. 2001) and perhaps
SGRs represent alternative pathways of young neutron stars,
distinguished by the history of mass inflow ($\dot{M}$) from a fall-back
accretion disk. This work classified young neutron stars according to 
ranges of $\dot{M}$, taken as representative constant values,
with radio pulsars corresponding to zero or very weak $\dot{M}$, such that
the disk does not quench the pulsar magnetosphere. The inner
radius of the disk must lie at or beyond the light cylinder. Disks around
radio pulsars were first proposed by Michel \& Dessler (1981, 1983) and 
Michel (1988).  For the AXPs Chatterjee, Hernquist \& Narayan (2000) 
studied a fall-back disk with a specific time dependent 
mass-inflow rate $\dot{M}(t)$ taken to follow a
self-similar thin disk solution, which entails a power-law decay of
$\dot{M}(t)$ (Cannizzo, Lee \& Goodman 1990, Mineshige, Nomoto \& 
Shigeyama 1993).
According to this model some neutron stars go through propeller 
and accretion (AXP) phases as $\dot{M}(t)$ evolves while others, 
at magnetic fields $<3 \times 10^{12} G$ become radio pulsars after a 
very brief initial accretion phase.

What are the implications of a fall-back disk
for radio pulsars? Marsden, Lingenfelter and
Rothschild (2001a,b) argued that the presence of a propeller spin-down
torque from a fall-back disk in conjunction with the magnetic dipole
radiation torque may explain the large discrepancy (Gaensler \& Frail
2000) between the real (kinematic) age and the "characteristic" age 
$P/2\dot{P}$
corresponding to pure dipole spin-down for the pulsar B1757-24, and that 
pulsar\--SNR age discrepancies can be explained in a similar way. Menou,
Perna and Hernquist (2001b) showed that this combination of dipole and
propeller torques can
explain the braking indices $n<3$ of five young pulsars.

In this letter we apply the combined spin-down torque, using the 
model of Menou, Perna and Hernquist (2001b), to the distribution of   
pulsars in the $P-\dot{P}$ diagram. 
Presence of torques other than dipole radiation torques,
in particular torques that might arise from the ambient medium near a
young pulsar in a supernova remnant was proposed first by Yusifov et al. (1995) 
to explain qualitatively the measured braking indices $n<3$ of young pulsars, 
the discrepancy between real and characteristic ages and the distribution of 
pulsars in the $P-\dot{P}$ diagram. Gvaramadze (2001) invokes torques from 
circumstellar clumps in the SNR to derive a "true" age for PSR B1509-58 
in agreement with the SNR age.
In Section 2, we explore the combined torque model, using the
canonical magnetic dipole torque together with the model of Menou,
Perna and Hernquist (2001b) for the propeller spin-down torque, with a 
constant mass inflow rate $\dot{M}$. Evolutionary tracks in the $P-\dot{P}$ 
diagram are presented in Section 3 along with analytical expressions 
for timespans of various phases, braking indices and other properties of 
the tracks. The $P-\dot{P}$ diagram is divided into strips delineated by 
tracks of 
constant magnetic field and mass inflow rate. Each strip is divided into 
period bins and a histogram is constructed for the number of pulsars in each 
period bin. A curve for the expected number of 
pulsars as a function of period is calculated according to the model and 
compared with the histogram for each strip. The results are discussed in Section 4.

\section{Spin-Down of a Pulsar with a Fall-Back Disk}

We model spin-down of a   pulsar under the combined action of magnetic
dipole radiation and propeller spin-down torques as
\begin{equation}
I\Omega\dot{\Omega}=-\beta\Omega^4-\gamma
\end{equation}
adopting the model of Menou, Perna and Hernquist (2001b)
with their notation. The neutron star will continue to act
as a radio   pulsar as long as the fall-back disk does not protrude into the
light cylinder. If the disk were detached from the light cylinder, it
would not exert any torque on the neutron star and its magnetosphere.
Assuming that the disk is attached to the light
cylinder, the torque can be estimated as:
\begin{equation}
N_{disk}=-2\dot{M}r_{lc}^2[\Omega-\Omega_{K}(r_{lc})]
\cong-2\dot{M}r_{lc}^2\Omega =-2\dot{M}c^2/\Omega \equiv -\gamma/\Omega
\end{equation}
where $\dot{M}$ is the mass inflow rate interacting with the light cylinder
and being ejected from the disk; $r_{lc}=c/\Omega$ is the light cylinder
radius, and $r_{lc}^2\Omega$ is the specific angular momentum extracted
from the  pulsar magnetosphere, since the Keplerian rotation rate
$\Omega_K(r_{lc})$ in the disk is small compared to 
the rotation rate $\Omega$ of the neutron star 
and its magnetosphere. The parameter 
$\gamma=2\dot{M} c^2 = 2 \times 10^{31} \dot{M}_{10}\,$ erg/s
is the rate of energy loss of the neutron star 
due to the propeller torque; $\dot{M}_{10}$ is the mass inflow 
rate in units of $10^{10}$ gm/s. The rate of energy loss due to magnetic 
dipole radiation is given by
\begin{equation}
\dot{E}_{dipole}=-{B_{\bot}}^2 R^6 \Omega^4/6c^3=-\beta\Omega^4 .
\end{equation}
Here $B_\bot$ is the component of the dipole magnetic field at the 
neutron star surface in the direction perpendicular to the rotation axis 
and R is the neutron star radius. This defines
$\beta=6.17 \times 10^{27} {B_{\bot,12}}^2 {R_6}^6$ .
Menou, Perna and Hernquist (2001b) give the solution of Eq.(1) for 
constant $\dot{M}$ as 
\begin{eqnarray}
t\, & = & \,\tau\,[\,\arctan((\beta/\gamma)^{1/2}{\Omega_i}^2)\,-\,
\arctan((\beta/\gamma)^{1/2}\Omega(t)^2)\,]\,
\\ & = 
& \,\tau\,[\,\pi/2\,-\arctan((\beta/\gamma)^{1/2}\Omega(t)^2)\,]\nonumber
\end{eqnarray}
where $\Omega_i$, the initial rotation rate of the  pulsar, is always large
enough to justify the second equation. The timescale $\tau$ is
$\tau=I/2(\gamma\beta)^{1/2}= 4.5 \times 10^7 
{\dot{M}_{10}}^{-1/2}{B_{\bot,12}}^{-1}$ yrs. 
In the $P-\dot{P}$ diagram pulsars will follow tracks given by Equation (1),
\begin{equation}
\dot{P} = (4\pi^2\beta/I)\, P^{-1} + (\gamma/4\pi^2 I)\, P^3 .
\end{equation}
Menou, Perna and Hernquist (2001b) have used this model to explain the 
observed values of the braking index
$n=\ddot{\Omega}\Omega/{\dot{\Omega}}^2$.
From their model we derived ($\dot{M}$,$B_\bot$) values 
($2.8\times 10^{16}$ gm/s, $7.1\times 10^{12}$ G) for the Crab pulsar, 
($1.5\times 10^{16}$ gm/s, $8.9\times 10^{12}$ G) for PSR B0540-69, 
($1.4\times 10^{15}$ gm/s, $5.2\times 10^{12}$ G) for the Vela pulsar, 
($2.8\times 10^{13}$ gm/s, $8\times 10^{13}$ G) for PSR J1119-6127, 
and ($3.6\times 10^{14}$ gm/s, $3\times 10^{13}$ G) for PSR B1509-58. 
\footnote{The magnetic field of PSR J1119-6127 is $B_\bot$ = $8.2 \times 
10^{13}$ G if one assumes dipole spin-down. Camilo et al (2000) report 
$B_\bot = 4.1\times 10^{13}$ G in the discovery paper due to a 
missing factor of 1/2 in the definition of the neutron star's dipole 
moment.}
\section{The Distribution of Pulsars in the $P-\dot{P}$ Diagram}

We used $P$ and $\dot{P}$ data from the Princeton Pulsar Catalog (Taylor et 
al. 1996).  
Tracks corresponding to Eq.(5) are shown in Fig 1.
At early times (small $P$), the dipole term dominates and the   pulsar
follows the left branch of the track, essentially spinning down under
the magnetic dipole radiation torque. The
braking index for pure dipole spin-down is n=3, and the left branch of each 
track follows the slope $n^{\prime} = 2-n = -1$ in the $\log \dot{P}-\log P$ 
diagram, 
\begin{equation}
P\dot{P} = (4\pi^2\beta/I)\propto{B_\bot}^2.
\end{equation} 
The transition to the regime in which the propeller torque dominates occurs 
at the minimum of a track, at the period
\begin{equation}
P_0=(2\pi)(\beta/3\gamma)^{1/4}=2\pi/\Omega_0.
\end{equation}
The slope $n^{\prime}=0$, and the braking index $n=2$ at 
this point. The minimum value of the spin-down rate is found to be
\begin{equation}
\dot{P}_{min}=(4\pi^2\beta/I) P_0^{-1} + (\gamma/4\pi^2I) P_0^3
=(8\pi/I)\gamma^{1/4}(\beta/3)^{3/4}.
\end{equation}
The propeller spin-down torque is dominant along the right branch of each 
track. The asymptotic behaviour, reached already at periods of about
$2P_0$, has the slope $n^{\prime} = 3$ 
($n = -1$). The evolution is very rapid on this branch.
Neutron stars will spin-down to $\Omega = 0$  ($P = \infty$) in a finite 
time: 
\begin{equation}
t_{max} = \pi \tau/2 = 7 \times 10^7 {{\dot{M}}_{10}}^{-1/2} 
{B_{\bot,12}}^{-1} yrs. 
\end{equation}
A third of this time is spent until reaching the minimum $\dot{P}$ at $P_0$.
About $0.23 t_{max}$ is spent until reaching $n = 2.5$, $P =
(3/7)^{1/4} \cong 0.8P_0$, when the pulsar has started deviating from
dipole spin-down. The evolution slows down at periods around the turning
point at $P_0$ to $P = \surd 3P_0$, $n = 0$ . From $n = 2.5$, $P \cong 0.8
P_0$ to $P=\surd 3 P_0$, $n = 0$ it takes $0.44 t_{max}$. As the   pulsar
proceeds into the propeller spin-down branch, the evolution speeds up, as
$\dot{P} \propto P^3$; spin-down from $P = \surd 3 P_0$ to $P = \infty$
takes only $t_{max}/3$.

Pulsar activity will turn off at reaching a critical voltage (the "death
lines", "the death valley") at a period satisfying
$B_{\bot,12}/P^2 = \alpha =  0.1 - 1$ (see Fig 1). The propeller
spin-down tracks are parallel to the death lines.
In our analysis of the $P-\dot{P}$ diagram we find that the minimum mass 
inflow
rate for the pulsars in our sample is $\dot{M}_{min} = 4.5 \times 10^8$ gm
s$^{-1}$. The propeller spin-down track for $\dot{M}_{min}$ bounds
the pulsar population on the right and is of the
form $B_{\bot,12}/P^2 = 0.3$.
This track is already in the death valley. Pulsars with $\dot{M} < 
\dot{M}_{min}$
will reach the death valley and turn off while they are still on the
dipole spin-down track. Pulsars with $\dot{M} > \dot{M}_{min}$ will evolve on
their propeller spin-down tracks until they reach $P_{death} =
(B_{\bot,12}/\alpha)^{1/2}$ at an age
\begin{equation}
t_{death} = t_{max} [ 1 - (2/\pi)\arctan(0.7\alpha(\dot{M}_{10})^{-1/2})].
\end{equation}
This age for pulsar turnoff is close to $t_{max}$ for almost all cases.
Even for the track at $\dot{M}_{min}$ with $\alpha = 0.3$ we find $t_{death}
= 0.5 t_{max}$. At times close to $t_{max}$ evolution is rapid. Few pulsars
are observed at late times (long periods) along the propeller spin-down 
tracks.
Therefore pulsar turnoff will not effect the
distribution of pulsars in the $P-\dot{P}$ diagram except for the lowest 
$\dot{M}$.
The spindown lifetime $t_{max}$ is plausible in view of the low $\dot{M}$
values obtained for the pulsar tracks. A time dependent disk would reach
$\dot{M} \approx 10^8 -10^{10}$ gm s$^{-1}$on timescales $\sim 
10^{8}$ yrs. If the disk is depleted before the pulsar completes 
its evolution the 
pulsar will switch to the pure dipole spin-down track corresponding to its 
magnetic field.

The number $\Delta N$ of pulsars with magnetic field $B_{\bot}$ and 
mass inflow rate $\dot{M}$ in a period interval $\Delta P$ at period $P$ is
\begin{eqnarray}
\Delta N & = & (\Delta N(P;\beta,\gamma)/\Delta P) \Delta P  
= R  f(P, \dot{P}) (\Delta t(P;\beta,\gamma)/\Delta P) \Delta P 
\\ & = & R  f(P, \dot{P})(\dot{P}(P;\beta,\gamma))^{-1} \Delta P\nonumber
\end{eqnarray}
Here $R$ is the birthrate of pulsars in the galaxy, and f($P$, 
$\dot{P}$), the
fraction of the galactic disk volume in which pulsars of period $P$ and 
period derivative $\dot{P}$ are observable, describes selection effects. For
simplicity we take f($P$, $\dot{P}$) to be constant, assuming that the
variation in f($P$, $\dot{P}$) is negligible compared to the variation of
$\dot{P}$. We further assume that one and only one track ($\beta,\gamma$)
passes through each point in the $P-\dot{P}$ plane, such that the plane can 
be divided into strips separated by chosen ($\beta,\gamma$) tracks. 
For each strip we count the pulsars in equal sized period bins 
and construct the histogram of $\Delta N$. We then choose a 
representative track ($\bar{\beta},\bar{\gamma}$) for
the strip such that the function 
$\Delta N(P)$ = constant/$\dot{P}(P$;$\bar{\beta},\bar{\gamma})$ matches 
the histogram. 

The combined dipole and propeller spin-down model has a minimum $\dot{P}$ 
at the period $P_0$ on each track. 
$\Delta N\propto\dot{P}^{-1}$ must have a maximum
at $P_0$; this is
where a pulsar's motion across the $P-\dot{P}$ diagram is slowest.
The model predicts that $\Delta N \propto\dot{P}^{-1}$ will increase in
proportion to $P$ at short periods, and decrease in proportion to $P^{-3}$ at
$P > P_0$. The histograms do have this property. For each strip, we fit
the histogram by choosing $\bar{\beta}$ to have the average value for the
strip and choosing $\bar{\gamma}$ such that $P_0$ is in the period
bin with the maximum number $\Delta N_{max}$ of pulsars. The 
normalization of the model curve is chosen to match the histogram at 
($P_0$, $\Delta N_{max}$). Four strips separated by boundary tracks with
$B_{\bot,12} = 50$, $4$, $2$, $0.8$ and $0.1$ are shown on the 
$P-\dot{P}$ diagram in Fig.1. For the upper four boundary tracks $P_0 = 
0.3$ s, and for the lowest boundary track at $B_{\bot,12} = 0.1$, $P_0 = 
0.8$ s.

\section{Results and Discussion}

The number distribution of pulsars in the $P-\dot{P}$
plane is fitted remarkably well with the combined effect of dipole and
propeller spin-down. The mass inflow rates range from a few 10$^{16}$ gm/s
for the youngest pulsars to a minimum of $4.5\times 10^8$ gm/s. 
The histograms for the four strips shown in Fig.2 show reasonable agreement 
with representative model curves. 
The model curves representing the four strips are characterized by 
$B_{\bot,12}=35.5$, $P_0=0.15$ s (Fig.2a); $B_{\bot,12}=3.16$, $P_0=0.3$ s 
(Fig.2b); $B_{\bot,12}=1.52$, $P_0=0.3$ s (Fig.2c) and $B_{\bot,12}=0.57$, 
$P_0=0.8$ s (Fig.2d).
The agreement with the model is not 
sensitive to the choice of strip boundaries except for $B_{\bot,12} < 0.4$ 
where a second peak appears in the histograms at $P \simeq 1.3$ s, 
as seen in Fig.2d. 
This excess may reflect disk depletion resulting in pulsars dropping 
down, from 
many propeller spin-down tracks at higher $B_{\bot}$ and $\dot{M}$, to their 
corresponding pure dipole spin-down tracks at periods close to death lines. 

$P_0$ is found to vary between 0.15 s and 0.8 s. 
From Eq.(7), the limited
range of $P_0$ suggests a correlation 
between $B$ and $\dot{M}$. Such a correlation is not expected physically. It
arises because a model with constant $\dot{M}$ is used to represent the
real spindown which occurs under a time-dependent $\dot{M}(t)$. The
period $P_0$ reflects the value of $\dot{M}$ at time $t_0$, when
$\dot{P}$ is minimum. 
To understand the effect in a simple way, suppose at $P_0(t_0)$ the dipole
and propeller torques experienced by the pulsar are equal. Suppose the mass
inflow rate follows a power law decay $\dot{M}(t)\propto t^{-\eta}$.
The pulsar is initially evolving down its dipole
spindown track, with $P(t)\propto t^{1/2}$. 
We can define a
$P_0(t)=P_0(\beta,\dot{M}(t))$, the turning point into the propeller 
track, according to the current $\dot{M}(t)$. From Eq.(7), 
$P_0(t)=2\pi(\beta/3\gamma(t))^{1/4}\propto t^{\eta/4}$.
For early times $P(t)<P_0(t)$. One can imagine the
pulsar moving along its dipole track with $P(t)\propto t^{1/2}$ while the
junction with the propeller track at the current $\dot{M}(t)$ moves down
the dipole track, ahead of the pulsar, but at a slower rate 
$P_0(t)\propto t^{\eta/4}$. 
At $t=t_0$, $P(t_0)=P_0(t_0)$ and the pulsar 
is just turning into its current propeller track.
For $t>t_0$, the pulsar follows its
current propeller track while the track itself continues to shift
towards the bottom-right corner of the $P-\dot{P}$ diagram, as $\dot{M}(t)$
continues to decrease.
To derive the form of $\dot{M}(t)$
let us use the observation that $P_0(t_0)=P_0(\beta,\dot{M}(t_0))$ is 
similar for all pulsars.
Taking $P_0=0.3$ s gives $\dot{M}(t_0)=1.98\times10^{11}B_{\bot,12}^2$. 
Dipole spindown (Eq.(6)) until $t=t_0$ yields $P_0=1.23\times10^{-4} 
B_{\bot,12} t_0$(yr)$^{1/2}=0.3$ s or 
$B_{\bot,12}^2=5.95\times10^6/t_0$(yr). Thus we find that 
$\dot{M}(t)=1.18\times10^{18}$(gm/s)$t$(yr)$^{-1}$, that is, $\eta=1$. For 
$B_{\bot,12}=1$, 
$t_0=5.95\times10^{6}$ yr and $\dot{M}(t_0) \cong 2\times10^{11}$ gm/s
are obtained.

The application here to all pulsars accross the   $P-\dot{P}$
diagram and its success in fitting the pulsar distribution require and 
support
the presence of low mass, low $\dot{M}$ disks, which are active for 
pulsar   
lifetimes of the order of $10^7$ yrs and remain attached to the light  
cylinder.
The time dependence of the disk and the mass inflow rate it supplies could
be incorporated  in a more detailed calculation. The self similar
isolated thin disk models with a power law decay of $\dot{M}(t)$ were 
employed by Menou, Perna and Hernquist (2001b) who
point out that for the Crab pulsar, PSR B0540-09 and PSR B1509-58, thin disks
in the $\dot{M}$ ranges indicated by the braking indices of these pulsars are
consistent with observational constraints in the optical though most of
the disk luminosity would be in the UV where detections are much harder.
For the Vela pulsar only a very small and highly inclined
disk could be compatible with observational constraints. 
According to the classification of Alpar (2001) disks are present
around all classes of young neutron stars,
providing accretion for AXPs and acting as propellers on DTNs and RQNSs.
For two AXPs and the RQNS in Cas A luminosities predicted by thin disk models
are ruled out or tightly constrained by observations in the optical and IR
(Coe \& Pightling 1998, Hulleman et al 2000, Hulleman, van Kerkwijk
\& Kulkarni 2000, Kaplan,  Kulkarni \& Murray 2001). 
Further, it is likely that disks
under propeller boundary conditions are not standard thin disks (Alpar 2001).
As the mass inflow is stopped at the disk-magnetosphere
boundary and largely ejected from the system, the disk may be
enshrined in a corona and outflow of ejected matter, possibly with   
larger effective area and softer spectrum.
Menou, Perna
and Hernquist (2001a) have shown that the thin disks may become neutral
and stop evolving as thin disks. The transition would quench the
disk's luminosity, but 
it would also suppress the dynamical evolution of the disk if the 
 viscosity of MHD origin is no longer operational. However, a smaller but 
nonzero viscosity should be operating in a neutral disk. Furthermore 
irradiation by the pulsar will 
probably keep the disk ionized and allow the mass inflow to continue, 
possibly as a power law decay. The minimum rotational energy loss rate of 
the pulsars is  
of the order of $10^{30}$ erg s$^{-1}$. The luminosity of the 
irradiated disk may be shifted to different spectral bands (Perna and 
Hernquist 2000).

The present application to the $P-\dot{P}$ diagram shows no indication of
an effect on the pulsar distribution due to depletion of the disk before
pulsar turnoff, except possibly for the oldest pulsars, well advanced in
their evolution on propeller spin-down tracks, dropping down to pure
dipole tracks near death lines. This may indicate
that there is no separate disk timescale. Disks interacting with pulsars
via propeller torques may be driven by these torques to evolve on the
propeller spin-down timescales of the pulsars. These problems, as well as
the assumption that the disk remains attached to the light cylinder
require further work on the coupling between the disk and the pulsar
magnetosphere. A Monte-Carlo simulation of the $P-\dot{P}$ diagram will
test the present model with independent random distributions of $B$
and $\dot{M}$ and selection effects.                                     

We predict that some pulsars have braking indices $<2$, like the Vela pulsar
(though the $n=1.4\pm 0.2$ braking index measurement for this pulsar should
be treated with some caution as its timing behaviour is dominated by
interglitch relaxation). About 2/3 of all pulsars will be on the propeller
spin-down branch, with $n<2$, at $P>P_0$. Half of these pulsars will have
negative braking indices, $-1<n<0$. Measurement of braking indices
is unfortunately very difficult since the timing behaviour is
dominated by noise (Baykal et al 1999) and it seems also by
interglitch recovery (Alpar and Baykal 2001). 

The model also 
predicts very small numbers $\Delta N \propto P^{-3}$ of the oldest 
pulsars, which have long 
periods and large $\dot{P}$ values. These pulsars would give 
magnetic fields $B_{\bot} \sim 10^{12}-10^{13}$, perhaps $10^{14}$ G 
with the combined
dipole-propeller spin-down model. For these pulsars the pure dipole 
spin-down model would yield higher fields, extending into the magnetar 
range, 
and young ages. Thus kinematic age measurements of pulsars wih long 
periods and large $\dot{P}$ can distinguish between pure dipole spindown and 
propeller spindown.
Propeller spin-down also increases the rate of energy dissipation in
pulsars and in neutron stars evolving under propeller torques after pulsar
activity is over.

\acknowledgments
We thank O. H. Guseinov and \"{U}. Ertan for discussions and S. 
\c{C}. \.{I}nam for technical support.
We thank T\"{U}B\.{I}TAK, the Scientific and Technical Research Council of 
Turkey, for  
support through TBAG-\c{C}G4 and through the BDP program for doctoral 
research. AA \& EY thank T\"{U}B\.{I}TAK 
for graduate student scholarships. MAA thanks the Turkish Academy of 
Sciences and Sabanc\i\ University for research support.

\newpage 

\noindent REFERENCES

\noindent  
Alpar, M.A. 1999, http://xxx.lanl.gov/abs/astro-ph/9912228\newline
Alpar, M.A. 2001, http://xxx.lanl.gov/abs/astro-ph/0005211, ApJ,
in the press\newline
Alpar, M.A. \& Baykal, A. 2001, to be submitted\newline
Baykal, A., Alpar, M.A., Boynton, P.E. \& Deeter, J.E. 1999, MNRAS, 306, 207\newline
Camilo, F. et al. 2000, ApJ, 541, 367\newline
Canizzo,J.K., Lee,H.M. \& Goodman,J. 1990, ApJ, 351,38\newline
Chakrabarty,D. et al. 2001, ApJ, 548, 800\newline
Chatterjee,P., Hernquist, L. \& Narayan, R. 2000, ApJ, 534, 373\newline
Coe,M.J. \& Pightling,S.L. 1998, MNRAS, 299, 233\newline
Gaensler, B.M. \& Frail, D.A. 2000, Nature, 406, 158\newline
Gvaramadze, V.V. 2001, http://xxx.lanl.gov/abs/astro-ph/0102431\newline
Hulleman, F., van Kerkwijk, M.H., Verbunt, F.W.M. \& Kulkarni, S.R. 2000, A\&A,
358, 605\newline
Hulleman, F., van Kerkwijk, M.H. \& Kulkarni, S.R. 2000,
Nature, 408, 689\newline
Illarionov,A.F. \& Sunyaev,R.A. 1975, A\&A, 39, 185\newline
Kaplan, D.L., Kulkarni, S.R. \& Murray, S.S. 2001,
http://xxx.lanl.gov/abs/astro-ph/0102054\newline
Marsden, D., Lingenfelter, R.E. \& Rothschild, R.E. 2001a, ApJ 547, L45\newline
Marsden, D., Lingenfelter, R.E. \& Rothschild, R.E. 2001b,
http://xxx.lanl.gov/abs/astro-ph/0102049\newline
Menou, K., Perna, R. \& Hernquist, L. 2001a,
http://xxx.lanl.gov/abs/astro-ph/0102478\newline
Menou, K., Perna, R. \& Hernquist, L. 2001b,
http://xxx.lanl.gov/abs/astro-ph/0103326\newline
Mereghetti, S. 1999, to appear in The Neutron Star - Blackhole Connection,
C. Kouveliotou, J. van Paradijs \& J. Ventura, eds, (Kluwer: Dordrecht),
http://xxx.lanl.gov/abs/astro-ph/9911252\newline
Michel, F.C. 1988, Nature, 333, 644\newline
Michel, F.C. \& Dessler, A.J. 1981, ApJ, 251, 654\newline
Michel, F.C. \& Dessler, A.J. 1983, Nature, 303, 48\newline
Mineshige, S., Nomoto, K. \& Shigeyama, T. 1993, A\&A, 267, 95\newline
Perna, R. \& Hernquist, L. ApJ 2000, 544, L57\newline
Taylor, J.H., Manchester, R.N. \& Lyne, A.G., 1996, The Princeton Pulsar 
Catalog, http:\\pulsar.ucolick.org/cog/pulsars/catalog\newline
Thompson, C. \& Duncan, R.C. 1995, MNRAS 275, 255\newline
Treves, A. et al. 2000, PASP, 112, 297\newline
Woods, P.M. et al. 1999, ApJ 519, L139\newline
Yusifov, I.M., Alpar, M.A., G\"{o}k, F., \& Guseinov, O.H. 1995, M.A. Alpar, 
\"{U}. K\i z\i lo\u{g}lu, J. van Paradijs (eds.) The Lives of the Neutron 
Stars, P.201, Dordrecht: Kluwer.

\newpage

\noindent Figure Captions

Figure 1. The log $P$(s)$ - $log$\dot{P}$ (s s$^{-1}$) diagram. Strips 
used for the histograms in Fig.2 are separated by the tracks shown,
$B_{\bot,12}$ = 50 - 4 (Fig.2a), $B_{\bot,12}$ = 4 - 2 (Fig.2b), 
$B_{\bot,12}$
= 2 - 0.8 (Fig.2c), and
$B_{\bot,12}$ = 0.8 - 0.1 (Fig.2d). The dashed lines are death lines
$B_{\bot,12}/P^2 = \alpha$ for $\alpha$ = 1 , 0.3 and 0.1 (left to right).

Figure 2. Histograms showing the number of pulsars in period bins for 
each of
the strips shown in Fig.1. The model curve on each histogram is drawn for the
average $\beta \propto B_{\bot,12}^2$ for the strip,
with $\gamma \propto \dot{M}$ chosen such that the maximum of the histogram 
occurs at $P_0 = 2\pi (\beta/3\gamma)^{1/4}$.

\clearpage

\begin{figure*}
\plotone{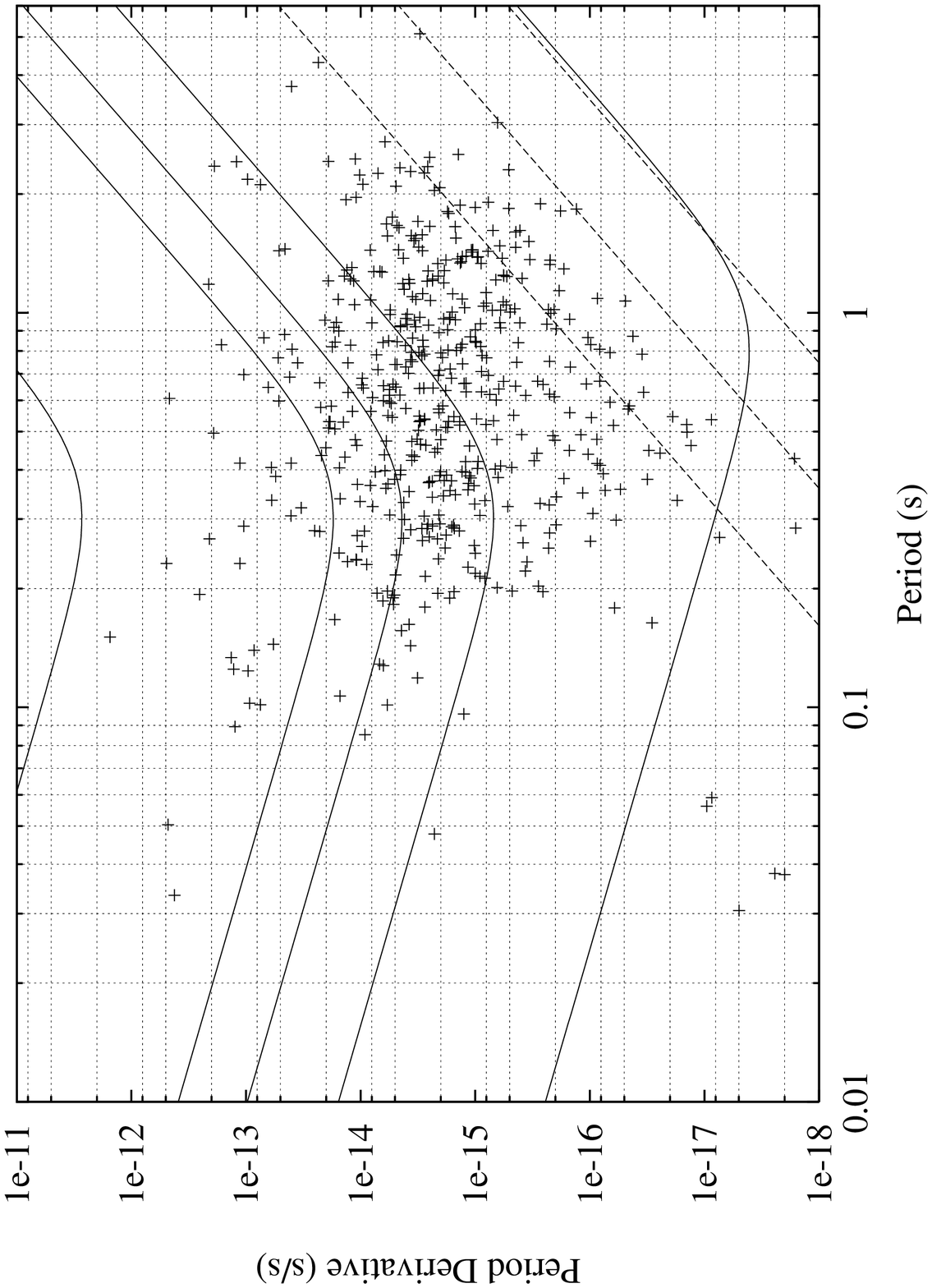}
{Fig.1}
\end{figure*}

\clearpage

\begin{figure*}
\plotone{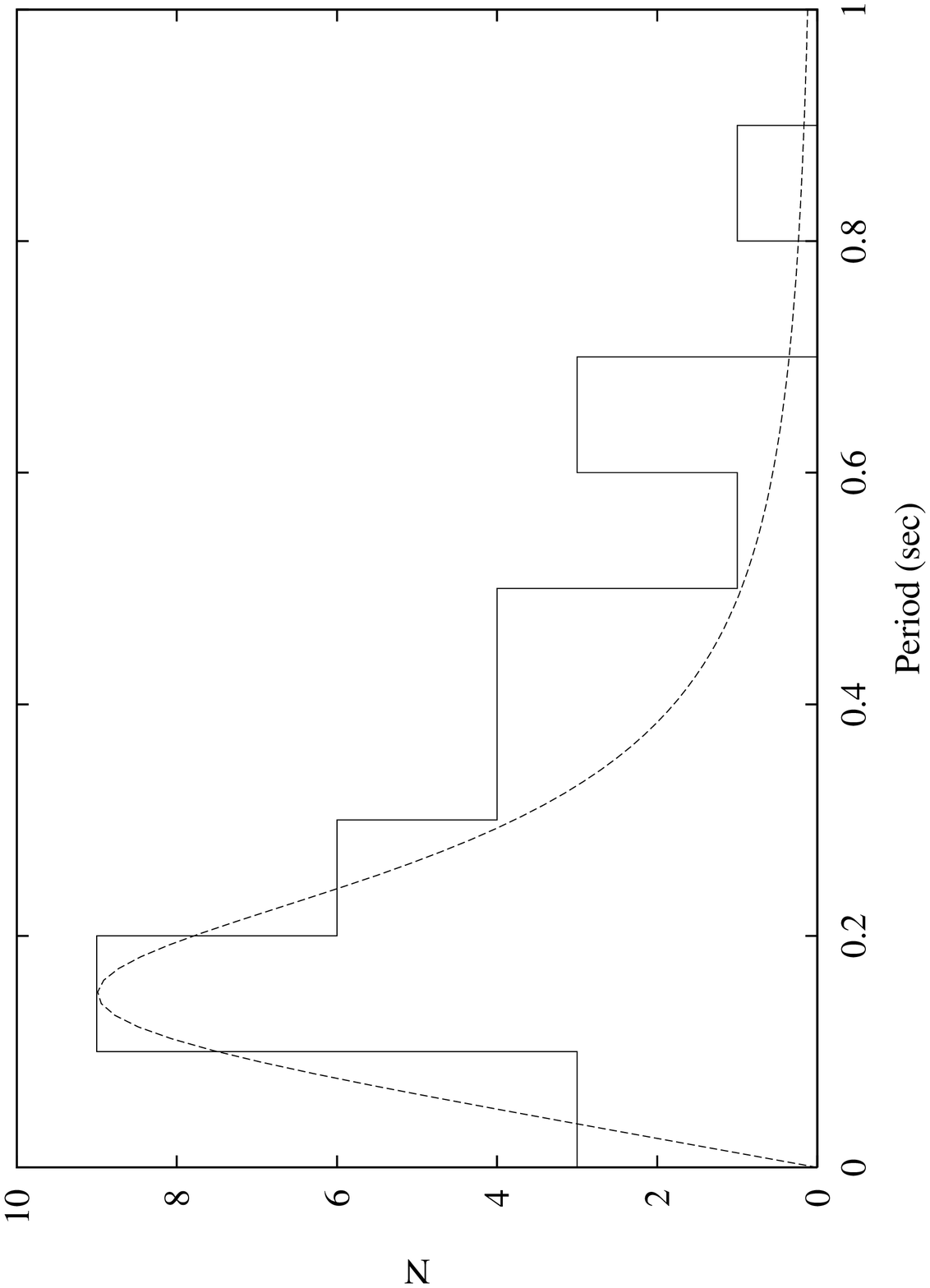}
{Fig.2a}
\end{figure*}

\clearpage

\begin{figure*}
\plotone{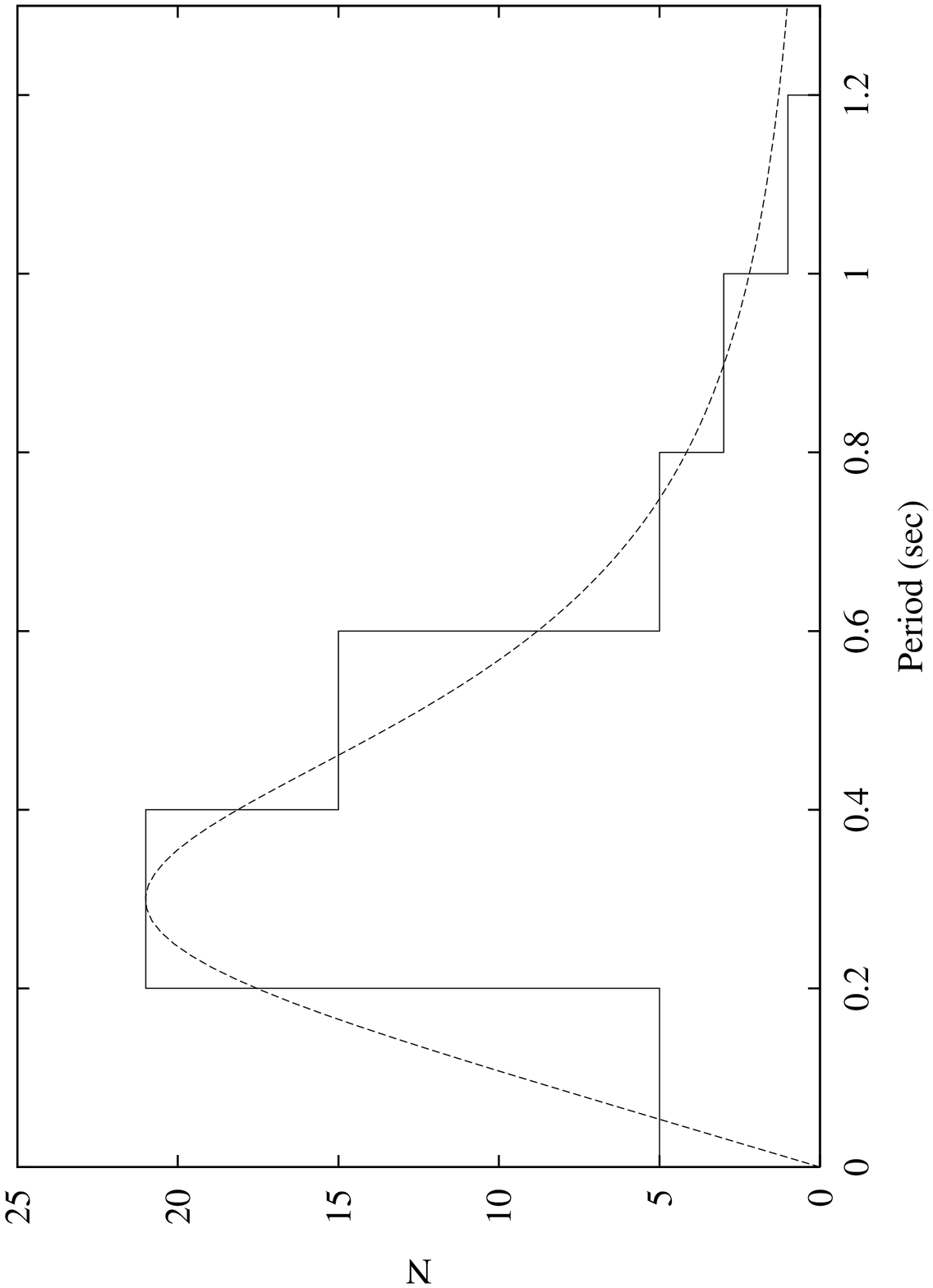}
{Fig.2b}
\end{figure*}

\clearpage

\begin{figure*}
\plotone{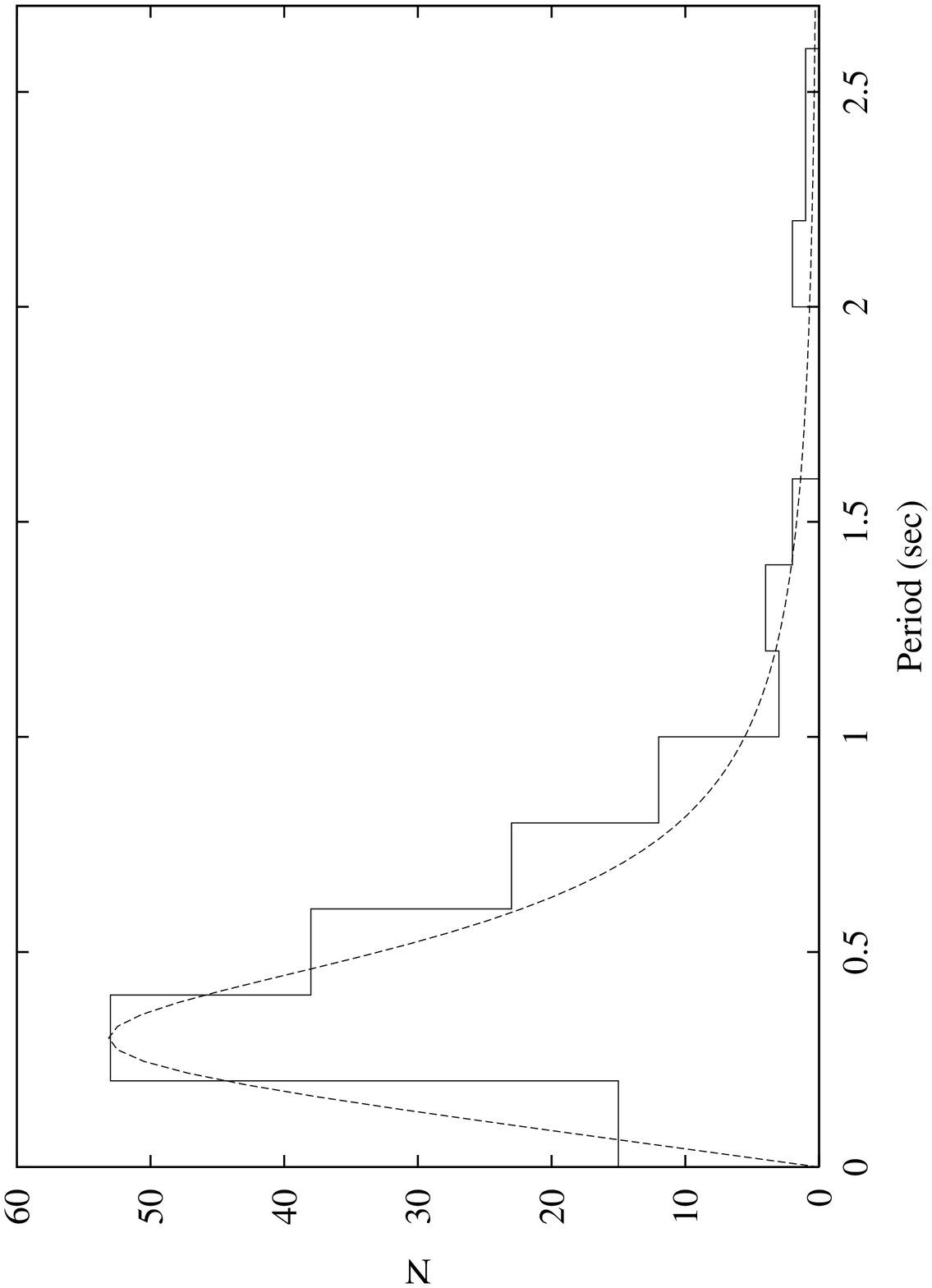}
{Fig.2c}
\end{figure*}

\clearpage

\begin{figure*}
\plotone{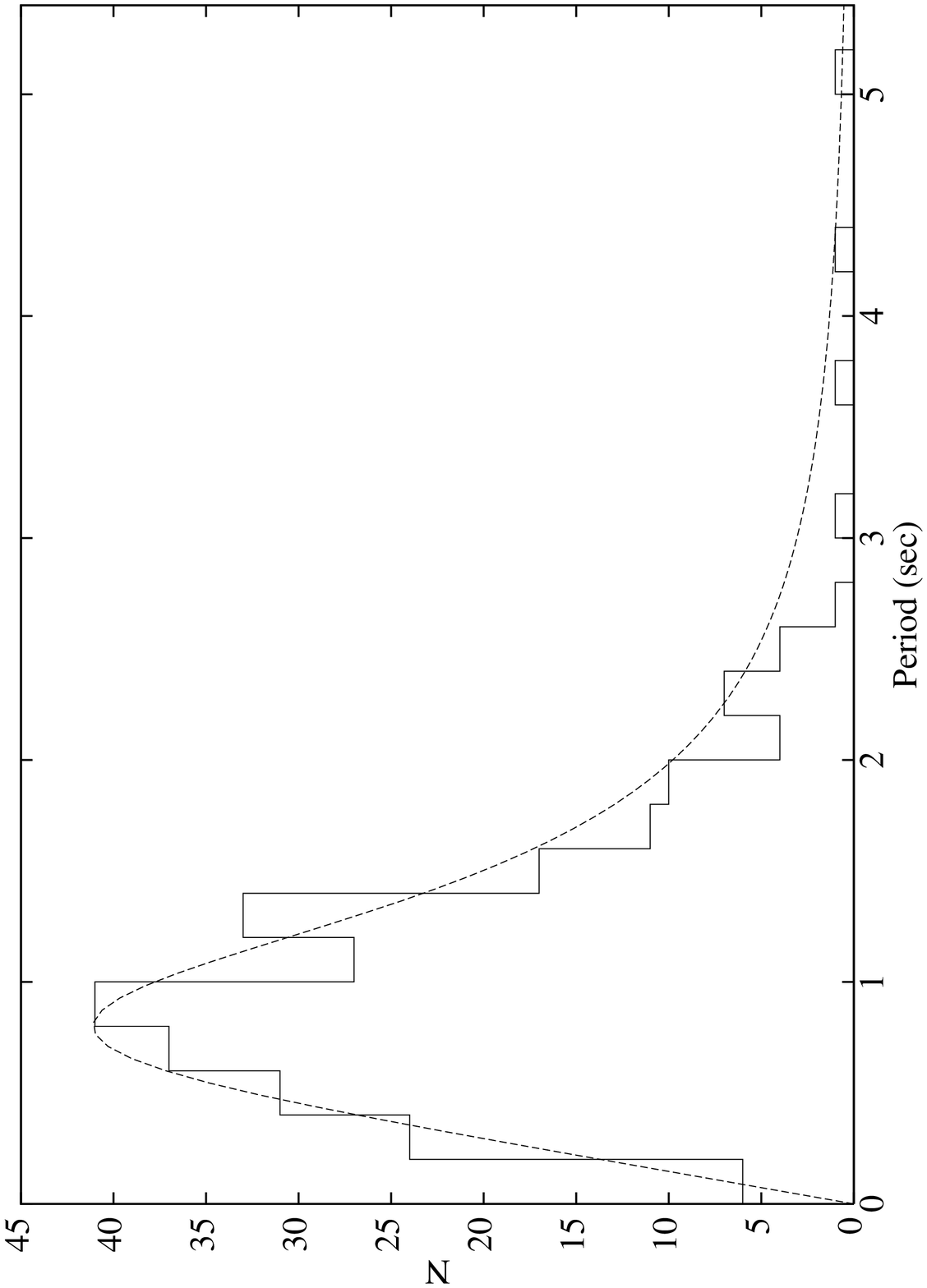}
{Fig.2d}
\end{figure*}

\end{document}